\documentclass[aps,preprint,showpacs,preprintnumbers,amsmath,amssymb,floatfix]{revtex4}
\usepackage{graphicx}
\begin{document}

\title{Formation of dynamical Schr\"odinger cats
in low-dimensional ultracold attractive Bose gases}

\author{Alexej I. Streltsov, Ofir E. Alon, and Lorenz S. Cederbaum}

\affiliation{Theoretische Chemie, Physikalisch-Chemisches Institut, Universit\"at Heidelberg,\\
Im Neuenheimer Feld 229, D-69120 Heidelberg, Germany}
\begin{abstract}
Dynamical Schr\"odinger cats can be formed when
a one-dimensional attractive Bose-gas cloud
is scattered off a potential barrier.
Once formed, these objects are stable in time.
The phenomenon and its mechanism --
transformation of kinetic energy to internal energy of the scattered atomic cloud --
are obtained by solving the 
time-dependent many-boson Schr\"odinger equation. 
Implications are discussed.
\end{abstract}

\maketitle
Famous Schr\"odinger cat-states (cats) \cite{ES} 
are subject of numerous studies 
in connection to different branches of quantum physics,
touching fundamental problems of quantum mechanics and 
theories of informations \cite{ex1,ex2,ex3,ex4}.
However, 
practical realization of these states,
in particular for larger systems,
is always an experimental challenge because 
these fragile systems are very sensitive 
to internal interactions, 
experimental noise and 
other imperfections.
In the context of Bose gases several
propositions have been made to creating 
Schr\"odinger cats with larger number of particles \cite{s1,s2,s3,s4,s5,s6}.
Here we report on a completely different physical
mechanism that leads to {\it dynamical Schr\"odinger cats} 
which can be accessible within present experimental setups in
low-dimensional Bose gases -- {\it by scattering an attractive 
Bose-gas cloud off a potential barrier}.
Once formed, these objects are stable in time.

Dynamics of low-dimensional Bose gases with attractive 
interparticle interaction has drawn much attention \cite{BS1,BS2,e1,e2,t1,t2,t3,frag}.
The ground state of one-dimensional (1D) bosons interacting via 
attractive delta-function interaction in free space is analytically known: 
it is given by a Bethe ansatz solution for any finite 
number of $N$ bosons \cite{Mc} 
and, 
for large number of particles, 
it approaches the famous bright-soliton solution of the Gross-Pitaevskii mean-field theory \cite{Cal}.
Many-body excited states are also accessible \cite{Mc} , however, 
with increasing $N$ it becomes not easy to operate 
with a large number of excited states needed for dynamical studies.
Most importantly,
for the general {\it nonuniform} case there are no known analytical solutions and 
therefore numerical quantum propagation remains the only available tool.
We use the multiconfigurational time-dependent Hartree method for bosons (MCTDHB) \cite{MCTDHB} to attack the problem. 
In the MCTDHB($M$) method 
the time-dependent many-body wavefunction $\Psi(t)$
is expanded by all time-dependent permanents generated by distributing 
the $N$ bosons over $M$ time-dependent orbitals.
The orbitals as well as the time-dependent expansion coefficients 
are determined according to Dirac-Frenkel 
time-dependent variational principle \cite{MCTDHB}.

\begin{figure}
\includegraphics[width=9.6cm,angle=-90]{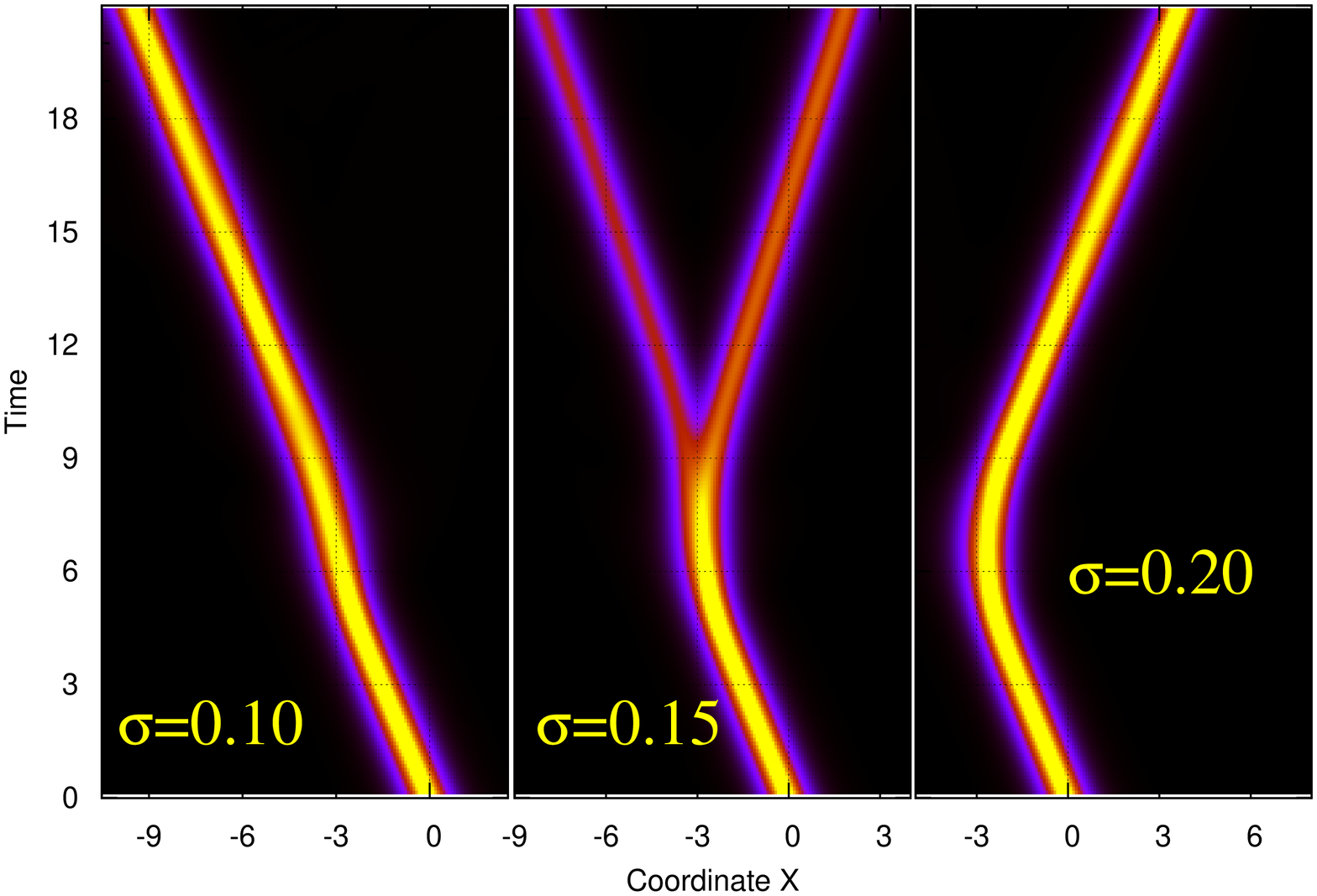}
\caption {(color online). 
Scattering of a solitonic wavepacket initially-located at $x=0$ and moving with constant velocity
off Gaussian barriers of different widths $\sigma$ centered at $x=-3$.
Shown is the density as a function of time.
Left panel: full transmission case.
Right panel: full reflection case.
Middle panel: formation of a dynamical Schr\"odinger cat in real space.
See text for more details.
All quantities are dimensionless.
}
\label{fig1}
\end{figure}

In the present work the initial state is an attractive cloud made of $N=100$ bosons,
prepared in the ground-state and localized around the origin.
To scatter it from a potential barrier we add some initial velocity to the cloud.
We use a Gaussian barrier placed quite apart from the initial wavepacket.
The time-dependent many-boson Schr\"odinger equation is solved for a fixed barrier 
height and several different barrier widths.
The results of these scattering ``experiments''
are plotted in Fig.~\ref{fig1};
the width of the barrier increases from left to right panels.
It is clearly seen that the wavepacket dynamics 
very much depends on the barrier width. 
In the left panel the wavepacket overcomes the barrier and continues to propagate 
further forward without dispersion and losses.
In the right panel the wavepacket is reflected from the barrier 
and continues to propagate backward without dispersion and losses.
The middle panel shows fundamentally different and unexpected dynamics: 
the initial wavepacket is split into two (unequal) parts. 
One of them is transmitted through the barrier 
and another one is reflected by the barrier.
After their formation,
each of the parts has its own constant 
velocity and continues 
to propagate without dispersion.
It will be shown below that 
this split object is
a {\it dynamical Schr\"odinger cat}. 

We recall that in the standard textbook-problem of scattering a single-particle 
wavepacket from a barrier there are generally transmitted {\it and} reflected waves.
It is thus anticipated that, 
due to the mutual attraction between the particles, 
a bright soliton being scattered off a barrier 
is either totally transmitted or totally reflected, 
depending on barrier parameters and initial kinetic energy.
The left and right panels of 
Fig.~\ref{fig1} represent this expected behavior,
while in the middle scenario this picture is violated. 
To understand this phenomenon, 
we begin with 
the energetics of the simulated scattering processes.
Throughout this work, 
we use dimensionless units for length, time and energy,
which are readily arrived at by dividing the 
Hamiltonian by $\frac{\hbar^2}{mL^2}$,
where $m$ is the mass of a boson and $L$ is a convenient length scale, say the size of the atomic cloud. 
The one-body Hamiltonian then reads: $\hat h(x)=-\frac{1}{2}\frac{\partial^2}{\partial x^2} 
+ V_0\exp\left[(x+3)^2/2\sigma^2\right]$. 
The height of the Gaussian barrier is set to $V_0=0.4$.
We employ the commonly-used contact interparticle interaction
$\lambda_0\delta(x-x')$ where $\lambda_0=-0.04$.
The energy of the cloud is $E_{\mathrm{GS}}/N=-0.66$.
The velocity added to every boson in the cloud is $v=-0.5$
[technically, this is achieved when every orbital in the initial 
many-body wavefunction $\Psi(0)$ is multiplied by the prefactor $\exp(+ivx)$],
resulting in kinetic energy of $T_{\mathrm {kin}}/N=v^2/2=0.125$.
Thus,
the total energy of the wavepackets $E_{\mathrm{GS}}+T_{\mathrm{kin}}$
and the barrier height $V_0$ are the same for the three presented simulations.
The only difference between them is 
the barrier width $\sigma$ used. 
In the first, full transmission case $\sigma=0.10$,
in the second scenario where the initial packet is split $\sigma=0.15$,
and in the third, full-reflection case $\sigma=0.20$.
So, the initial state is the same for all three scenario
and the barriers' topology and parameters are also quite 
close to each other.
Therefore, 
the nature of excitations available for the dynamics, i.e., 
induced by the barriers, 
should also be quite similar in the three cases as well.
Are they?

To answer the above posed question, 
we first have to analyze and understand the quantum nature of the propagated wavepackets. 
Here,
a few words on the many-body method used are in place. 
The key idea of MCTDHB
is to use a many-body ansatz for $\Psi(t)$
where both the expansion coefficients 
and one-particle functions are allowed to 
change in time and in an optimal manner \cite{MCTDHB}.
Such an ansatz admits global changes of the physical nature of the evolving wavepacket 
as well as local changes of each individual physical state involved.
Let us explain this. 
Evolution of the expansion coefficients 
can be comprehended as possible dynamical changes due to transitions
between quantum states of different physical nature, e.g., 
between condensed and fragmented states.
Time evolutions of the one-particle functions in real space can be considered as 
local dynamical changes of the density of a given quantum state, 
e.g., dipole, breathing, etc., excitations of a condensed state.
The famous Gross-Pitaevskii theory, 
which is usually used to describe dynamics of bright solitons, 
is a very particular case of MCTDHB theory,
where the global changes of the wavefunction are forbidden, i.e., 
the system is forced to stay condensed 
and only local dynamical changes of the density are allowed. 

\begin{figure}[ht]
\includegraphics[width=9.6cm,angle=-90]{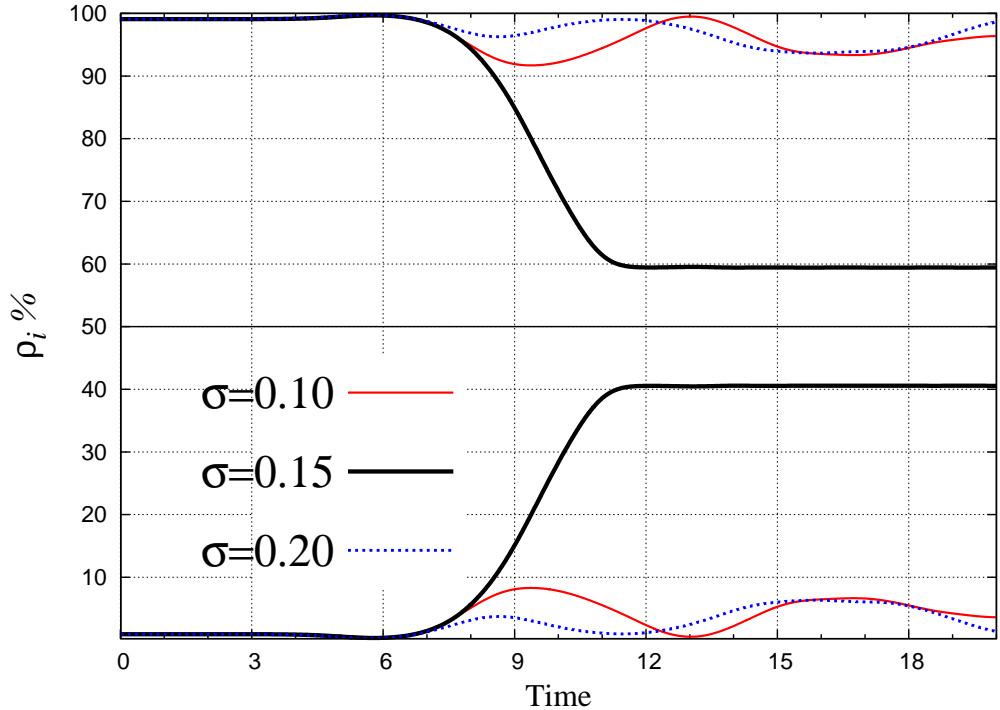}
\caption [kdv]{(color online). 
Natural occupation numbers $\rho_i(t)$ in \%
during the scattering processes of Fig.~\ref{fig1}.
At $t=0$ all initial wavepackets are the same and condensed: $\rho_1(0)=99.1\%,\rho_2(0)=0.9\%$.
Until $t=3$ the systems propagate without dispersion, 
reflecting the solitonic character of the initial wavepacket.
In the full reflection ($\sigma=0.10$) and transmission ($\sigma=0.20$) cases, 
the systems remain mainly condensed all the time.
In the split case ($\sigma=0.15$), 
after interaction with the barrier, 
the system evolves into a two-fold fragmented
state with essentially time-independent occupation 
numbers $\rho_1=59.5\%,\rho_2=40.5\%$. 
All quantities are dimensionless. 
}
\label{fig2}
\end{figure}

As a first step, 
we diagonalize at each point in time the 
reduced one-body density matrix
$\rho(x|x';t)=\sum_i\rho_i(t)\phi^\ast_i(x',t)\phi_i(x,t)$
for the three scenario,
where the eigenfunctions $\phi_i(x,t)$ are the natural orbitals.
In Fig.~\ref{fig2} we plot the obtained eigenvalues 
(natural occupation numbers) $\rho_i(t)$ as a function of time.
These quantities are very useful for a state characterization because according to standard definitions,
the system is {\it condensed} \cite{Penrose} if only one natural orbital is macroscopically 
occupied and {\it fragmented} \cite{Nozieres} if several natural orbitals have large eigenvalues, 
i.e., are macroscopically occupied.
Accordingly, the initial solitonic wavepacket is condensed, 
because almost all the bosons reside 
in one natural orbital, $\rho_1(0)=99.1\%$.
Moreover, 
till about $t=3$ 
the systems evolve without changes of the occupation numbers, see Fig.~\ref{fig1},
which indicates that indeed the propagating state is a solitonic wavepacket, 
moving with constant velocity and without dispersion.
Changes appear when the localized cloud starts to climb up the potential barrier.
In the transmission and reflection cases,
the interaction of the cloud with the barrier results in a small
redistribution of the occupation numbers
during and after the collision with the barrier.
We conclude that in the full transmission and reflection cases 
the system remains mainly condensed. 
In the interaction region, 
kinetic energy is transformed to potential energy 
when the cloud is climbing up the barrier and 
transfered back when the system is sliding down.
In other words, 
the internal state of the cloud is 
only slightly affected during the evolutions.
In the split case, however, 
the situation is very different, 
as one can see from Fig.~\ref{fig2}.  
During the interaction of the wavepacket with the barrier, 
$\rho_2(t)$ grows until it reaches some macroscopic value and saturates around it for large propagation times 
when the cloud is split.
We conclude that in the split case, 
due to interaction of the atomic cloud with the barrier, 
the system becomes two-fold fragmented and stays fragmented afterwords. 
In this case the initial kinetic energy is transformed 
to potential energy 
and globally changes 
the internal state of the system. 

Let us pose for a moment and summarize our main findings so far.
It has been shown that scattering a 1D attractive Bose gas in its ground-state off a potential barrier
can lead to a formation of a two-fold fragmented state.
The fragmented state is dynamically stable and retains its properties in time.
The mechanism involves transformation of kinetic energy to internal energy
of the scattered atomic cloud due to interaction with the barrier. 
What is the nature of this fragmented state in the attractive Bose system? 

\begin{figure}[ht]
\includegraphics[width=8.6cm,angle=-90]{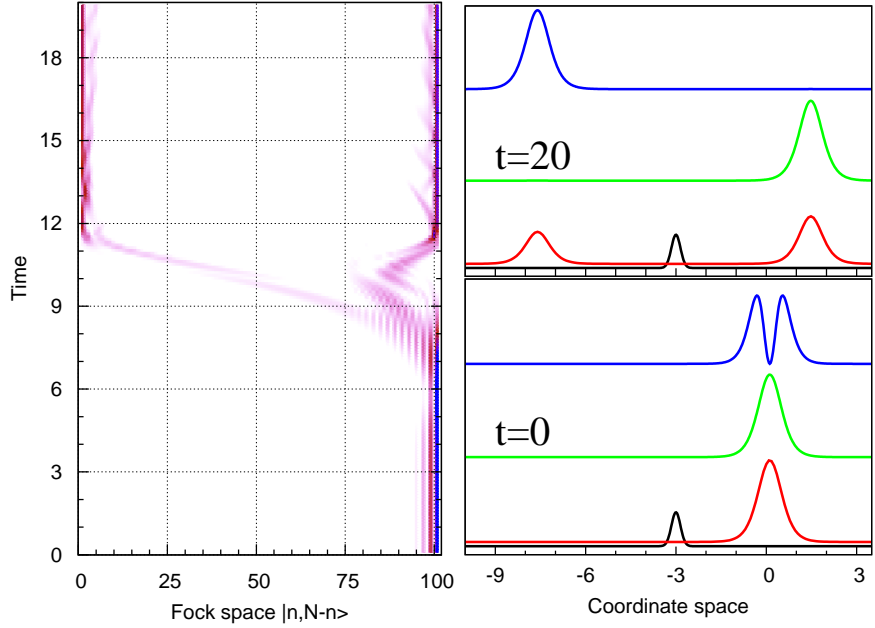}
\caption [kdv] {(color online). 
Proof that the split object is a Schr\"odinger cat.
Left panel: evolution of the expansion coefficients in Fock space
spanned by the $|N,0\rangle,|N-1,1\rangle,\ldots,|1,N-1\rangle,|0,N\rangle$ 
configurations. The probabilities $|C_n(t)|^2$ are plotted as a function of time.
The initial wavepacket is described essentially by $|0,N\rangle$.
Right panels: natural orbitals $|\phi_i(x,t)|^2$ (in green and blue; normalized to 1) and densities (in red) 
at $t=0$ and $t=20$.
The barrier is also shown.
The split object is a Schr\"odinger cat 
because mainly the $|N,0\rangle$ and $|0,N\rangle$ 
configurations contribute. 
All quantities are dimensionless.
}
\label{fig3}
\end{figure}

To get a deeper insight into the physics of this fragmented--split case, 
we investigate the many-body structure of the evolving wavepacket in more details. 
In the present study the total many-body wavefunction reads:
$|\Psi(t)\rangle =\sum_{n=0}^N C_n(t) {|n,N-n;t\rangle}$ where 
$\langle x_1,\ldots,x_N|n,N-n;t\rangle=
\hat{\cal S}\phi_1(x_1,t)\cdots\phi_1(x_n,t)\phi_2(x_{n+1},t)\cdots\phi_2(x_N,t)$
and $\hat{\cal S}$ the symmetrization operator.
This ansatz makes the condensed and all possible 
two-fold fragmented states available for the dynamics.
In what follows, 
we for brevity do not
indicate explicitly the 
time-dependence of the configurations.
To define the many-body wavefunction 
at a given time means 
to specify the (natural) orbitals $\phi_1(x,t),\phi_2(x,t)$ and 
respective expansion coefficients $C_n(t)$. 
This allows one for a graphical representation and analysis of the whole many-body wavefunction.
In Fig.~\ref{fig3} we analyze the fragmented--split case.
The natural orbitals in coordinate space
before ($t=0$) and after ($t=20$) 
the scattering process are plotted in the right panels of Fig.~\ref{fig3}.
The left part of Fig.~\ref{fig3} depicts 
the evolution in time of the corresponding expansion coefficients in Fock space,
spanned by the $|N,0\rangle,|N-1,1\rangle,\ldots,|1,N-1\rangle,|0,N \rangle$ permanents. 
For convenience, the time-dependent probabilities $|C_n(t)|^2$ are plotted.

First we discuss the real-space dynamics of the many-body wavefunction, i.e., the behavior of natural orbitals.
As seen in Fig.~\ref{fig3}, 
initially both natural orbitals are localized 
around the origin.
The first natural orbital ($\rho_1(0)=99.1\%$) has no nodes and 
the {\it marginally-occupied} second natural orbital ($\rho_2(0)=0.9\%$) has one node.
After the collision with the barrier and the split, at, e.g., $t=20$, 
both natural orbitals are localized and have a very similar one-hump--no-node shape, 
see Fig.~\ref{fig3}.
It is also important to notice that their profiles (widths) resemble 
the shape of the primarily-occupied 
natural orbital of the initial state.

Now we analyze the dynamics of the respective probabilities $|C_n(t)|^2$ in Fock space.
As shown in Fig.~\ref{fig3}, 
the initial state is described by a very narrow distribution 
of the expansion coefficients with maximal 
contribution provided by the single configuration $|N,0\rangle$.
This picture remains unchanged until $t=3$, 
where the solitonic many-body wavefunction starts to interact with the barrier.
From Fig.~\ref{fig3} we also see that, 
during this process,
more and more permanents become involved in the dynamics,
which is reflected in drastic changes to the overall pattern in Fock space. 
After the splitting, 
the pattern of the distribution of the expansion coefficients again becomes relatively simple --  
there are mainly two 
dominant configurations, $|N,0\rangle$ and $|0,N\rangle$, 
augmented by small contributions 
from a few respective neighboring states.

Let us now combine both observations. 
The initial wavepacket 
is mainly described by the configuration $|N,0\rangle$,
whereas the split object is formed by 
a superposition of mainly the $|N,0\rangle$ and $|0,N\rangle$ configurations.
The shapes of the occupied orbitals in {\it both} cases are very similar and,
most importantly, for the split object 
the orbitals are localized in different regions of space. 
All in all,
we can conclude that the split object is 
a dynamical realization of the famous Schr\"odinger cat.

Several remarks are in order.
First, the most important observation is 
that the split object is formed, 
and once formed it is stable in time.
Second, this split object is not a ``perfect'' Schr\"odinger cat. 
One can see from Fig.~\ref{fig1} 
that the density of the left split part is smaller than the density of the right part.
The quantitative characterization of this ``asymmetry'' 
can be obtained from Fig.~\ref{fig2}, 
where we see that the natural occupation numbers 
saturate at $59.5\%$ and $40.5\%$, 
accounting thereby for $59.5\%$ (right part) 
and $40.5\%$ (left part) 
of the total density, 
respectively.
Finally, 
our numerical calculations show that
a variety of dynamical Schr\"odinger cats can be obtained,
depending on the barrier shape, particle number and interparticle interaction strength.

It is instructive to contrast the Schr\"odinger cats reported here
with the fragmented states reported previously in \cite{frag}.
The fragmenton is another 
dynamically-stable fragmented object which is 
described essentially
by a {\it single} Fock configuration $|N/2,N/2\rangle$ build upon {\it delocalized} orbitals \cite{frag}. 
In contrast, 
the two-fold-fragmented Schr\"odinger cat is, 
by the definition, 
a {\it two-configurational} many-body state, i.e, 
it is formed as 
a linear combination of 
the two configurations $|N,0\rangle$ and $|0,N\rangle$ build upon {\it localized orbitals}.
Clearly, Schr\"odinger cats and fragmentons are of different but complimentary physical nature, 
that can be explained in terms of localization and delocalization in the real and Fock spaces. 
The fragmentons appear due to delocalization of the orbitals in real space and localization of 
configurations in the Fock space, 
while Schr\"odinger cats are formed due to delocalization 
of the expansion coefficients in the Fock space
and localization of the orbitals in real space.
An important physical distinction between these
two classes of dynamically-stable excitations in attractive Bose gases
is their energetics.
Schr\"odinger cats lie 
much lower in energy than fragmentons \cite{frag},
as is also reflected in the finding that their orbitals are essentially of the same shape
as the ground-state orbital (see Fig.~\ref{fig3}).

Let us conclude. 
An initial solitonic wavepacket being scattered from a potential barrier 
can form dynamically-stable Schr\"odinger cats. 
The initial kinetic energy is transformed due to interaction with the barrier 
to potential and internal energies,
resulting in formation of two not-exact replicas of the original wavepacket, 
augmented by small excitations of the local densities.
The condensed state is dynamically transformed to a fragmented state.
The implications of the present results go beyond the context of
Schr\"odinger cats with mesoscopic number of particles,
and demonstrate the wealth of low-dimensional attractive Bose gases
as ``hosts'' of novel many-body phenomena.
We hope that our work will stimulate experiments.

\begin{acknowledgments}
Financial support by DFG is acknowledged.
\end{acknowledgments}

\newpage

\end{document}